\documentclass[a4paper,11pt]{article}
\usepackage{pos}
\usepackage{xcolor}
\definecolor{themecolor}{RGB}{56, 92, 180} 
\definecolor{oiBlue}{RGB}{0,114,178}    
\definecolor{oiOrange}{RGB}{230,159,0}  
\definecolor{oiSkyBlue}{RGB}{86,180,233}
\definecolor{oiBluishGreen}{RGB}{0,158,115} 
\definecolor{oiYellow}{RGB}{240,228,66} 
\definecolor{oiVermillion}{RGB}{213,94,0}   
\definecolor{oiReddishPurple}{RGB}{204,121,167} 

\definecolor{alert}{RGB}{180, 0, 0}
\definecolor{example}{RGB}{0, 110, 0}

\title{Heavy quark masses from step-scaling}

\author*[c]{Simon Kuberski}
\author[mz,g]{Alessandro Conigli}
\author[d]{Patrick Fritzsch}
\author[mar]{Antoine G\'erardin}
\author[mu]{Jochen Heitger}
\author[ma]{Gregorio Herdo\'{\i}za}
\author[ma]{Carlos Pena}
\author[n]{Hubert Simma}
\author[n,h]{Rainer Sommer}

\affiliation[c]{Theoretical Physics Department, CERN, 1211 Geneva 23, Switzerland}

\affiliation[mz]{Helmholtz Institute Mainz, Johannes Gutenberg University, Mainz, Germany}
\affiliation[g]{GSI Helmholtz Centre for Heavy Ion Research, Darmstadt, Germany}

\affiliation[d]{School of Mathematics, Trinity College Dublin, Dublin 2, Ireland}

\affiliation[mar]{Aix-Marseille Universit\'e, Universit\'e de Toulon, CNRS, CPT, Marseille, France
}
\affiliation[mu]{Universit\"at M\"unster, Institut f\"ur Theoretische Physik,\\
	Wilhelm-Klemm-Stra{\ss}e 9, 48149 M\"unster, Germany}

\affiliation[ma]{Instituto de F\'{\i}sica Te\'orica UAM-CSIC and Dpto. de F\'{\i}sica Te\'orica\\
	C/~Nicol\'as Cabrera 13-15, Universidad Aut\'onoma de Madrid, Cantoblanco E-28049 Madrid, Spain}
	
\affiliation[n]{John von Neumann-Institut f{\"u}r Computing NIC, Deutsches Elektronen-Synchrotron DESY,\\
	Platanenallee 6, 15738 Zeuthen, Germany}
\affiliation[h]{Institut f{\"u}r Physik, Humboldt-Universit{\"a}t zu Berlin\\
	Newtonstr. 15, 12489 Berlin, Germany}

\emailAdd{simon.kuberski@cern.ch}

\abstract{
We present a determination of the charm- and bottom-quark masses using the heavy-quark step-scaling strategy. 
Renormalization is performed in small volumes where relativistic bottom quarks can be simulated directly.
A sequence of finite-volume simulations connects this calculation to large-volume CLS ensembles, where simulations at physical light and strange quark masses provide reliable control over low-energy hadronic physics.
In all but the smallest volume, the B-scale is reached by interpolating between relativistic heavy-quark data and the static limit.
The resulting quark masses are obtained with good precision, with subdominant systematic uncertainties that differ from, and thus complement, those of standard large-volume determinations.

\vspace*{0.5cm}
\begin{flushright}
	CERN-TH-2026-068 \\
	IFT-UAM/CSIC-26-34 \\
	MS-TP-26-11 
\end{flushright}
}

\FullConference{The 42nd International Symposium on Lattice Field Theory (LATTICE2025)\\
2--8 November 2025\\
Tata Institute of Fundamental Research, Mumbai, India}

\begin{document}
\maketitle

\section{Introduction}

Precise determinations of the heavy-quark masses are an important
ingredient for many phenomenological applications.
The charm and bottom quark masses enter predictions for
Higgs decays, heavy-flavour observables and global fits of Standard
Model parameters.

Standard lattice calculations determine heavy-quark masses from
hadronic observables computed in large volumes.
However, relativistic simulations of the bottom quark on typical
lattice spacings are prohibited due to large discretization effects.
This motivates strategies that separate the heavy-quark scale from the
scale associated with low-energy hadronic physics.

Step-scaling realizes such a strategy. 
Heavy-quark observables are computed in a small reference volume where
relativistic bottom quarks can be simulated directly.
Finite-volume step-scaling functions are then used to connect this
regime to large volumes where simulations with physical light-quark
masses are available.

Most existing lattice determinations of the bottom-quark mass rely on
effective theories or extrapolations from lighter quark masses
(see the overview in~\cite{FlavourLatticeAveragingGroupFLAG:2024oxs}).
In the following we present our strategy for determining heavy-quark
masses using the step-scaling approach and summarize the current
status of the calculation.

\section{Step-scaling strategy for heavy-quark masses}
The determination of quark masses from Ward identities is
independent of finite-volume effects, apart from discretization
effects, since it relies on operator relations.
This allows the bottom quark mass to be determined in physically small
volumes.
Such volumes permit very fine lattice spacings, keeping discretization
effects under control and enabling a controlled continuum
extrapolation.
A reliable determination nevertheless requires that the bare quark
mass is tuned precisely to its physical value and thus contact to large volumes must be made. 

In conventional large-volume lattice QCD simulations the bare quark
masses are tuned by requiring that a hadron mass sensitive to the
quark mass reproduces its physical value.
These reference values are typically taken from the PDG~\cite{ParticleDataGroup:2024cfk} or from the masses defining isoQCD
according to the Edinburgh consensus recommended by
FLAG~\cite{FlavourLatticeAveragingGroupFLAG:2024oxs}.
To connect a small-volume determination of the quark mass to such a
large-volume hadronic input, the corresponding finite-volume effects
must therefore be computed.

The step-scaling approach provides a tool to connect the small- and
large-volume regimes of QCD.
It was used for heavy-quark physics in~\cite{Guagnelli:2002jd,deDivitiis:2003iy} and
later refined in~\cite{Guazzini:2007ja}, while also being a key element in the non-perturbative renormalization and matching of HQET developed 
in~\cite{Heitger:2001ch,Heitger:2003nj}.
A comprehensive description of our strategy for B-physics observables
can be found in~\cite{Sommer:2023gap}; here we summarize only the
aspects relevant to quark masses.
The central difficulty is that computing the finite-volume effect for
a B-meson mass would require evaluating the B-meson mass itself in
large volumes, which is precisely what we aim to avoid.

The solution is to evaluate heavy-light observables at heavy quark masses
lighter than the bottom quark mass, where cutoff effects are under control,
and combine these data with a
computation of the same observable in the static limit of Heavy Quark
Effective Theory (HQET).
This supplements the relativistic simulations used in conventional
heavy-quark mass extrapolations with information from the effective
theory.
The resulting dataset allows for a smooth and controlled interpolation
guided by HQET, 
as will demonstrated later in Figure~\ref{fig:interpolation}.
In the ideal case a precise determination in the static limit strongly
constrains the observable at the physical bottom quark mass and
provides a test of higher-order corrections in the heavy-quark
expansion.

A complication arises from the need for a non-perturbative
renormalization of the effective theory and a non-perturbative
matching to QCD.
These steps are required to avoid power divergences in the inverse
lattice spacing~\cite{Maiani:1991az} and to reach the precision needed
for phenomenological applications~\cite{Sommer:2010ic}.
The key observation underlying the step-scaling strategy is that the additive matching and renormalization contributions of HQET cancel in static energy differences~\cite{Sommer:2023gap}.

We therefore formulate the strategy directly in terms of
finite-volume mass differences.%
\footnote{By a finite-volume mass we denote a quantity $m_H(L)$ which approaches a (hadron) mass in the limit  $L\to\infty$.}
The resulting step-scaling functions encode the finite-volume
corrections required to connect the small- and large-volume regimes.
They are defined as	
\begin{align}\label{e:ssf}
	\sigma_m(L_1) = L_{\rm ref} [m_H(2L_1) - m_H(L_1)]\,,\qquad \rho_m(L_2) =  L_{\rm ref} [m_H - m_H(L_2)]\,.
\end{align}
Here $m_H$ denotes a heavy-light meson mass computed either in a finite
volume or in a volume that is effectively infinite for the observable
under consideration.
The reference scale $L_{\rm ref}$ renders the quantities dimensionless and we use $L_{\rm ref} = L_2 = 0.9988(63)\,{\rm fm}$.
The step-scaling functions can therefore be computed fully
non-perturbatively both with relativistic heavy quarks and in the
static theory.
Combining the two determinations allows us to interpolate the
step-scaling functions to the physical bottom quark mass.

The strategy of our calculation can be summarized as follows:
\begin{enumerate}
	\item Compute heavy-light meson masses in large volumes with relativistic heavy quarks lighter than the bottom quark and in the static theory.
	\item Perform the same calculations in the smaller volumes $L_1 \approx 0.5\,{\rm fm}$ and $L_2 = 2L_1 \approx 1.0\,{\rm fm}$ and determine the corresponding step-scaling functions.
	\item Compute the renormalized heavy-quark mass with relativistic
	quarks in the volume $L_1$, where discretization effects are
	controlled.
\end{enumerate}

Combining these ingredients allows us to determine the renormalization group invariant (RGI) heavy-quark mass $m_h^{\mathrm{RGI}}$ through
\begin{align}\label{Lmh_def}
	m_h^{\mathrm{RGI}}
	=&{}
	\left[
	\textcolor{black}{m_{\mathrm{H}}^{\rm phys}}
	- \frac{\textcolor{themecolor}{\rho_m(L_2)}}{L_{\rm ref}}
	-\frac{ \textcolor{example}{\sigma_m(L_1)}}{L_{\rm ref}}
	\right]
	\textcolor{alert}{\frac{
			m_{h}^{\rm RGI}
		}
		{
			m_{\rm H}(L_1)
		}
	}\,.
\end{align}
In this equation, $m_{\mathrm{H}}^{\rm phys}$ denotes the heavy-light meson mass used to tune the heavy-quark mass.
We discuss the specific choice in Section~\ref{s:results_ssf}.
While the main target of this work is the bottom quark mass, we also
compute the charm quark mass in order to obtain a result that is
largely uncorrelated with existing determinations.

\begin{figure}[t]
	\centering
	\begin{minipage}[t]{0.59\textwidth}
		\vspace{0pt}
		\includegraphics[width=\textwidth]{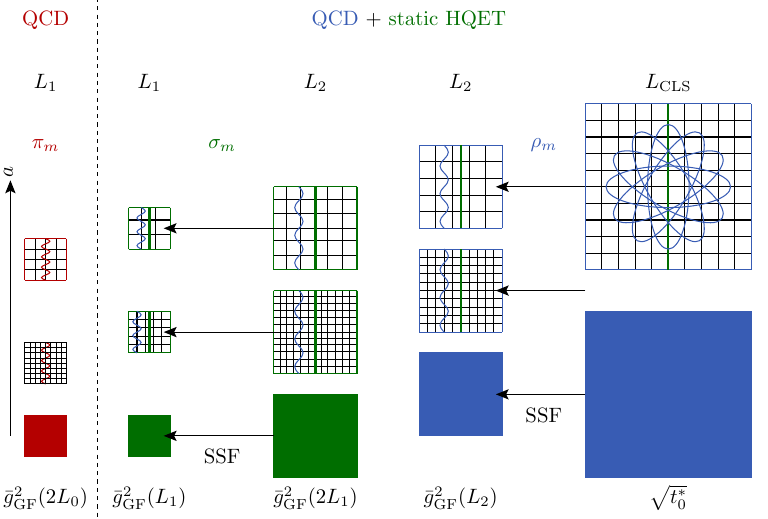}
	\end{minipage}\hfill
	\begin{minipage}[t]{0.39\textwidth}
		\vspace{0pt}
		\includegraphics[width=\textwidth]{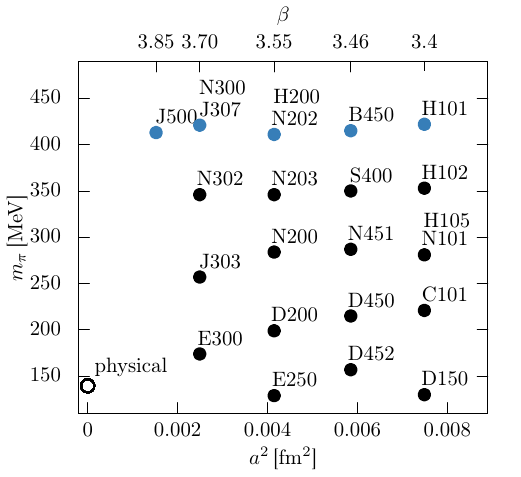}
	\end{minipage}
	\caption{
		\textit{Left:} Strategy for connecting small and large volumes via step-scaling functions, which are obtained from relativistic heavy-quark simulations and in the static limit and hence can be combined by interpolation.
		\textit{Right:} Overview of the CLS ensembles used in this work.
	}
	\label{fig:hqet_strategy}
\end{figure}

\section{Simulation setup}

The calculation combines simulations in a sequence of finite volumes
with large-volume CLS ensembles.
All simulations employ three flavors of $\mathrm{O}(a)$ improved Wilson
quarks~\cite{Bulava:2013cta} with the Lüscher--Weisz tree-level improved
gauge action.
The small-volume simulations are performed in the Schrödinger functional
(SF) with three massless quarks in volumes $V_i=L_i^4$, following the
setup of~\cite{DallaBrida:2016kgh}.
Three lines of constant physics are defined by vanishing sea quark masses
and fixed values of the gradient-flow coupling
$u_i=\bar g^2_{\mathrm{GF}}(L_i)$~\cite{DallaBrida:2016kgh},
as illustrated schematically in Figure~\ref{fig:hqet_strategy}.

Renormalization constants and improvement coefficients are computed
in a volume $L_0\approx0.25\,\mathrm{fm}$ defined by $u_0=3.949$.
Simulations are carried out at five lattice spacings down to
$a\approx0.0078\,\mathrm{fm}$, together with lattices with twice the
number of sites in each direction~\cite{Fritzsch:2018yag,Kuberski:2020rjq}.
These larger ensembles (red in Figure~\ref{fig:hqet_strategy}) are used to
determine renormalized heavy-quark masses with relativistic quarks.
A second set of simulations at volumes $L_1$ and $2L_1$ is used to
compute the step-scaling function $\sigma_m(L_1)$ with relativistic
and static heavy quarks.
In the final step the SF simulations at
$L_2=2L_1\approx1\,\mathrm{fm}$ are matched to the large-volume CLS
ensembles~\cite{Bruno:2014jqa} (blue in Figure~\ref{fig:hqet_strategy}).
Here, the SF simulations cover the same range of gauge couplings as the CLS
ensembles, allowing the two calculations to be connected.

To relate simulations with massless sea quarks to the large-volume
calculations with physical sea-quark masses we rewrite the
step-scaling function $\rho_m$ from Eq.~(\ref{e:ssf}) as
\begin{align}\label{e:ssf_rho}
	\textcolor{themecolor}{\rho_m(L_2)} =
	L_{\rm ref}\!\left[m_{\mathrm H}-m_{\mathrm H}^{\rm sym}\right]
	+L_{\rm ref}\!\left[m_{\mathrm H}^{\rm sym}-m_{\mathrm H}(L_2)\right]
	\equiv \tau_m(m_\pi,m_K)+\rho_m^{\rm sym}(L_2)\, .
\end{align}
Here $m_{\mathrm H}^{\rm sym}$ denotes the heavy-meson mass computed
with three mass-degenerate sea quarks at the flavor-symmetric point of QCD,
corresponding to $m_\pi=m_K\approx420\,\mathrm{MeV}$.

The right hand side of Figure~\ref{fig:hqet_strategy} shows the CLS ensembles used in this work.
They lie on the chiral trajectory where the sum of bare sea-quark
masses is kept constant.
The ensembles at the flavor-symmetric point entering the step-scaling function $\rho_m^{\rm sym}$
are highlighted in blue.

\section{Determination of the quark masses and the step-scaling functions \label{s:results_ssf}}

\begin{figure}[t]
	\centering
	\includegraphics[width=0.48\textwidth]{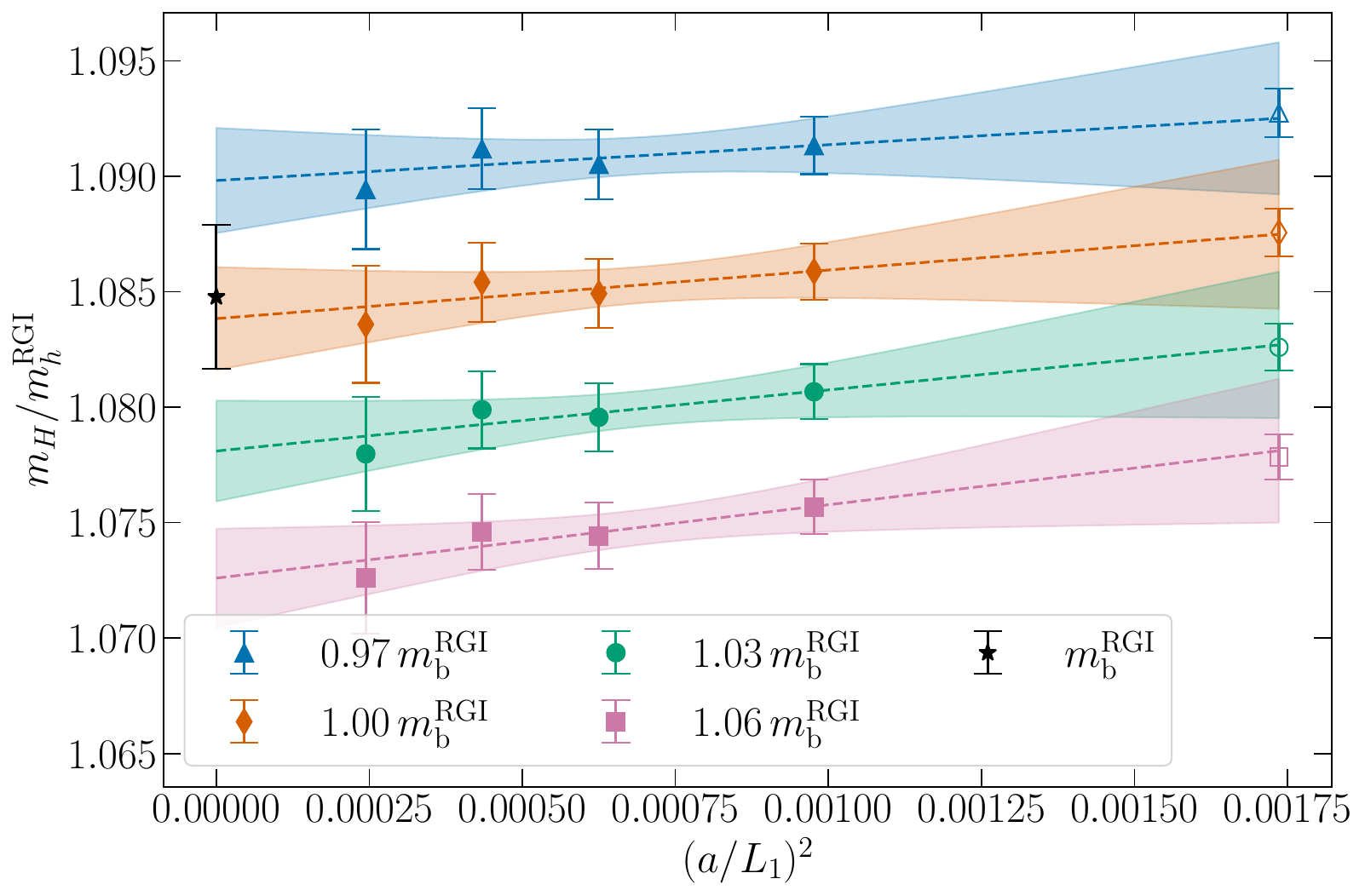}
	\includegraphics[width=0.48\textwidth]{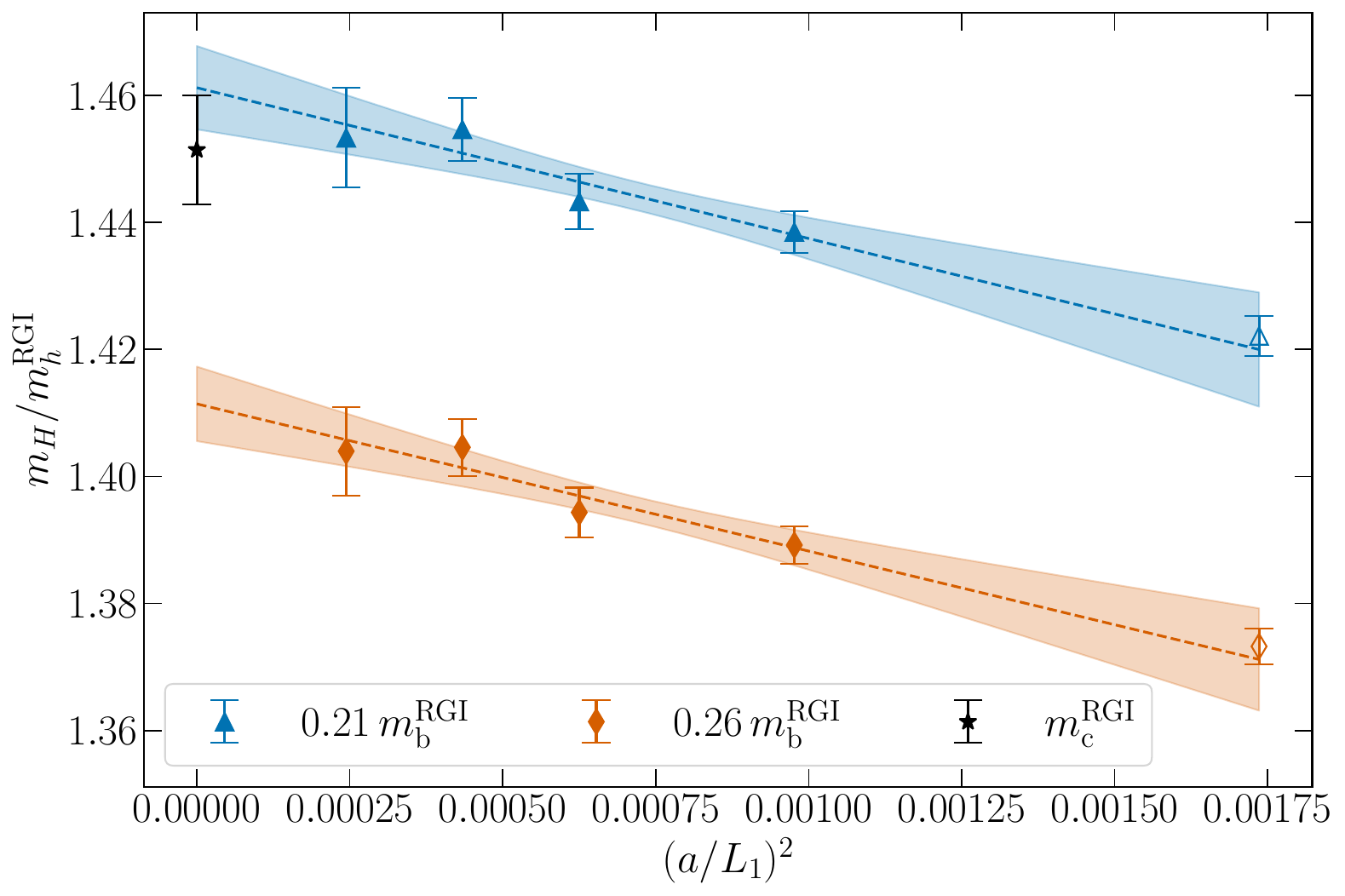}
	\caption{
		Continuum extrapolations of the ratio $m_H/m_h^{\rm RGI}$ in the
		volume $L_1$.
		The left panel shows masses around the bottom-quark region, while
		the right panel covers the charm region.
		Different colors correspond to different heavy-quark masses, and
		the black star in the continuum limit represents the interpolated
		result from the continuum-extrapolated values.
	}
	\label{fig:continuum_Q1s}
\end{figure}

We now present the determination of the step-scaling functions and the
resulting heavy-quark masses.
Figure~\ref{fig:continuum_Q1s} shows continuum extrapolations of the
ratio $m_H(2L_0)/m_h^{\rm RGI}$ for heavy-quark masses in
the bottom region (left) and the charm region (right).
Different colors denote different heavy-quark masses, while the black
point in the continuum limit is obtained by interpolating the
continuum-extrapolated values to the target finite-volume heavy-light meson mass.

The extrapolations are very smooth.
We estimate the associated systematic uncertainty by varying the cuts
in the lattice spacing and by changing the fit ansatz, taking the
anomalous dimension $\hat{\Gamma}$ in the lattice-spacing dependence
$(\alpha_s(1/a))^{\hat{\Gamma}} a^2$ to be either $0$ or
$0.76$~\cite{Husung:2021mfl}.

\begin{figure}  
	\centering
	\includegraphics[width=.48\textwidth]{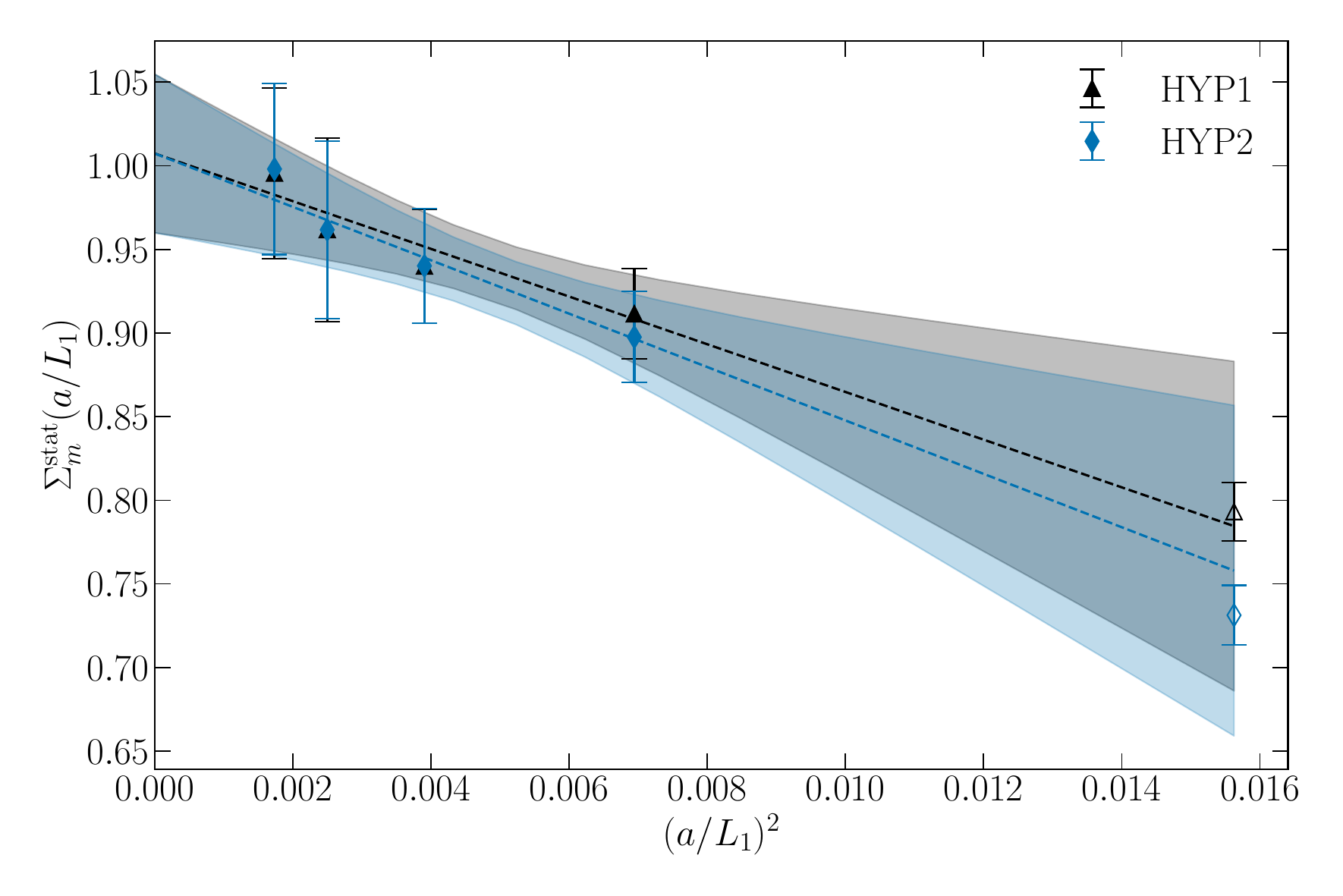}%
	\includegraphics[width=.48\textwidth]{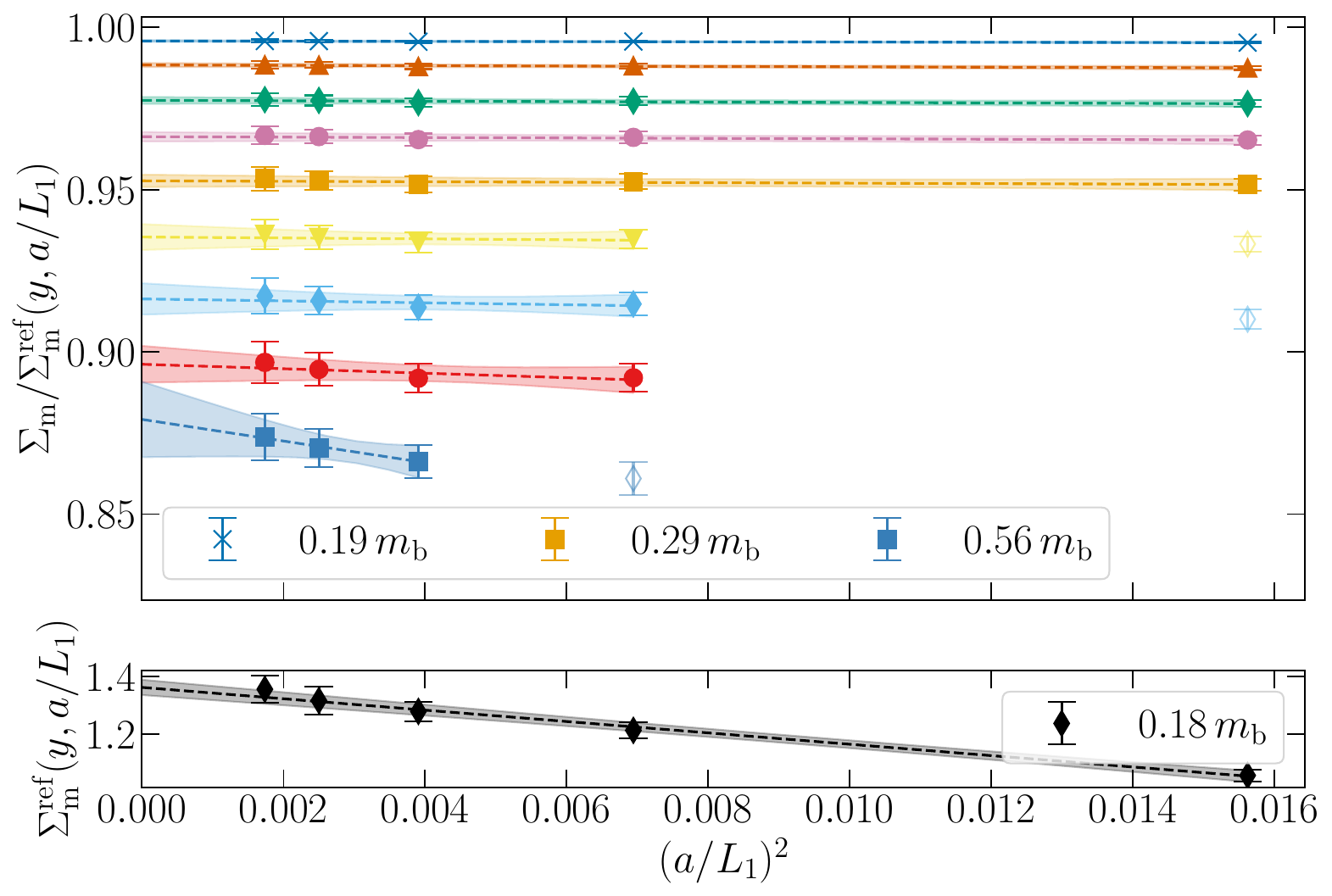}%
	\caption{
		Continuum extrapolations for the step-scaling function $\sigma_m$.
		Left: static limit using two discretizations of the static action.
		Right: relativistic heavy-quark results.
		The lower panel shows the continuum extrapolation near the charm
		mass, while the upper panel displays extrapolations of ratios of
		step-scaling functions at heavier masses relative to this
		reference mass.
	}
	\label{fig:continuum_H1}
\end{figure}

The continuum extrapolations entering the determination of the
step-scaling function $\sigma_m$ are shown in
Figure~\ref{fig:continuum_H1}.
In the static limit (left) we use two discretizations of the static
action based on HYP-smeared gauge fields~\cite{Hasenfratz:2001hp,DellaMorte:2003mn}, which approach a
common continuum limit.

For relativistic heavy quarks (right) the continuum limit is performed
in two steps.
The lower panel shows the extrapolation of $\sigma_m$ at a heavy-quark
mass close to the charm mass.
For heavier masses we instead consider ratios of step-scaling
functions with respect to this reference mass, whose continuum
extrapolations are displayed in the upper panel.
This strategy improves the control of discretization effects at large
heavy-quark masses, where we rely on the finer lattices only.

\begin{figure}[t]
	\centering
	\includegraphics[width=.48\textwidth]{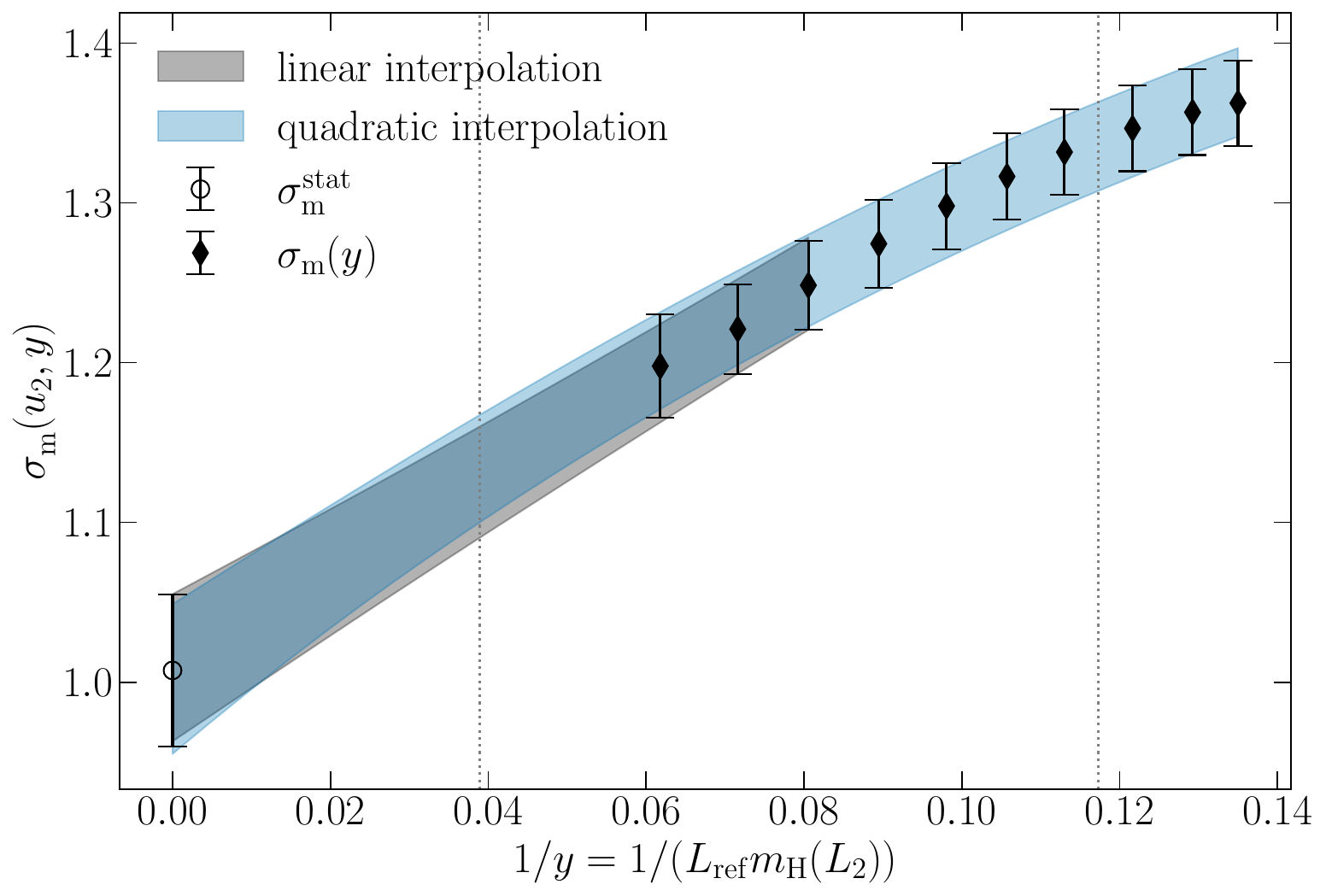}%
	\includegraphics[width=.48\textwidth]{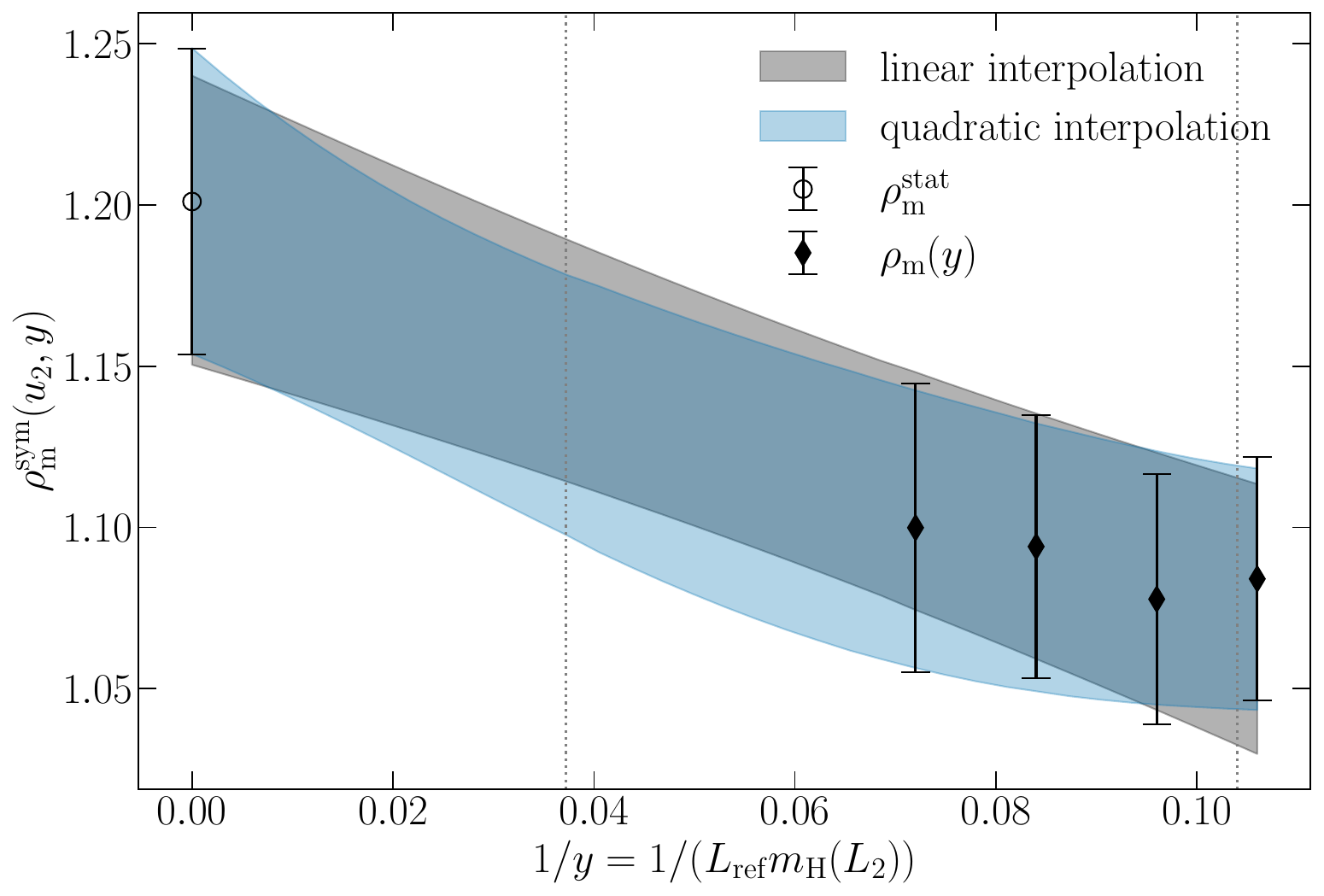}%
	\caption{
		Interpolation of heavy-quark observables between relativistic data
		and the static limit in the continuum limit.
		The left panel shows the step-scaling function $\sigma_m$ and the
		right panel $\rho_m^{\rm sym}$.
		Open circles denote the static-limit results, while black diamonds
		correspond to relativistic heavy-quark data.
		The gray band shows a linear interpolation in $1/y$ in the indicated
		region, whereas the blue curve corresponds to a quadratic
		interpolation using all data points.
		Vertical dotted lines indicate the target values for the bottom-
		and charm-quark masses.
	}
	\label{fig:interpolation}
\end{figure}

We now turn to the interpolation in the heavy-quark mass for the two
step-scaling functions $\sigma_m$ and $\rho_m^{\rm sym}$ in the
continuum limit, shown in Figure~\ref{fig:interpolation}.
The open circles denote the static-limit results, while the black
diamonds correspond to relativistic heavy-quark data.
To determine the observables at the charm- and bottom-quark masses we
interpolate the data as a function of
\begin{align}
	1/y = \frac{1}{L_{\rm ref}\, m_H(L_2)}\,,
\end{align}
which we use as a proxy for the inverse heavy-quark mass.
The gray band shows a linear interpolation in $1/y$ in the indicated
region, while the blue curve corresponds to a quadratic interpolation
using all available data.
The vertical dotted lines mark the charm- and bottom-quark masses.

The relativistic data cover different ranges in the two panels.
For $\rho_m^{\rm sym}$ (right) the step-scaling function involves
larger volumes and therefore larger lattice spacings.
To keep cutoff effects under control we restrict the relativistic
simulations to masses where $a m_h$ remains sufficiently small,
such that the data extend less far towards the static limit.
A small difference between the two interpolation ansätze is visible
and included in the systematic uncertainty.
Nevertheless the results remain fully consistent and the
uncertainties are well controlled, thanks to the point in the static limit,
which strongly constrains the interpolation at large heavy-quark
masses.

Lastly, we determine the dependence of the heavy-quark masses on the
light-quark masses in large volumes, encoded in the function
$\tau_m(m_\pi,m_K)$ introduced in Eq.~(\ref{e:ssf_rho}).
To minimize this effect we match to the flavor-averaged meson masses
$m_H^{\rm phys} = \overline{m}_{D} \equiv \tfrac{2}{3}m_D + \tfrac{1}{3} m_{D_{\rm s}}$
and
$m_H^{\rm phys} =  \overline{m}_{B} \equiv \tfrac{2}{3}m_B + \tfrac{1}{3} m_{B_{\rm s}}$.
Along our chiral trajectory these combinations are constant to 
leading order in chiral perturbation theory, implying that $\tau_m$ depends primarily on
$(m_K^2-m_\pi^2)^2$~\cite{QCDSF-UKQCD:2013tzk}.

\begin{figure}
	\includegraphics[height=.18\textheight]{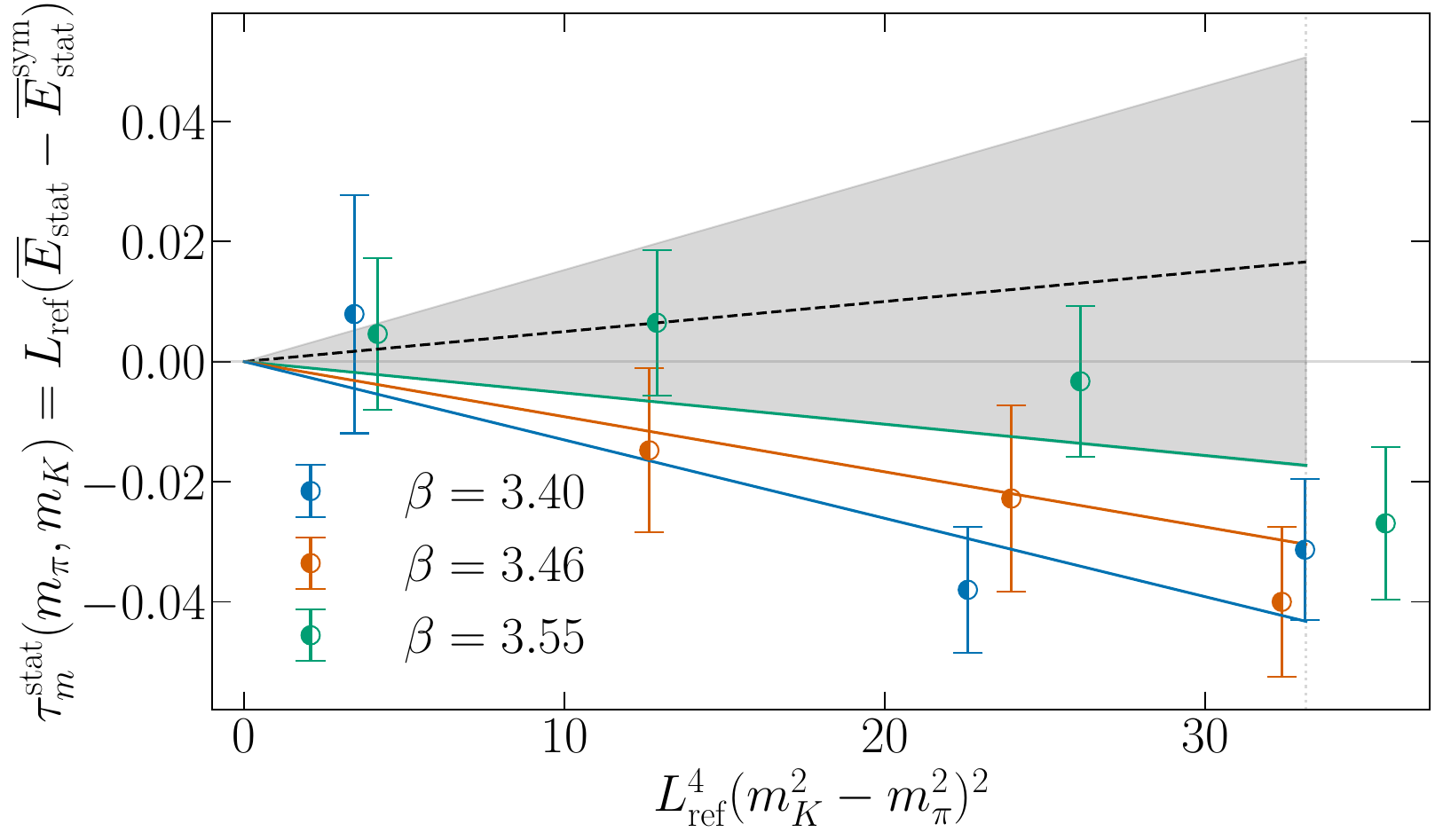}%
	\hfill
	\includegraphics[height=.18\textheight]{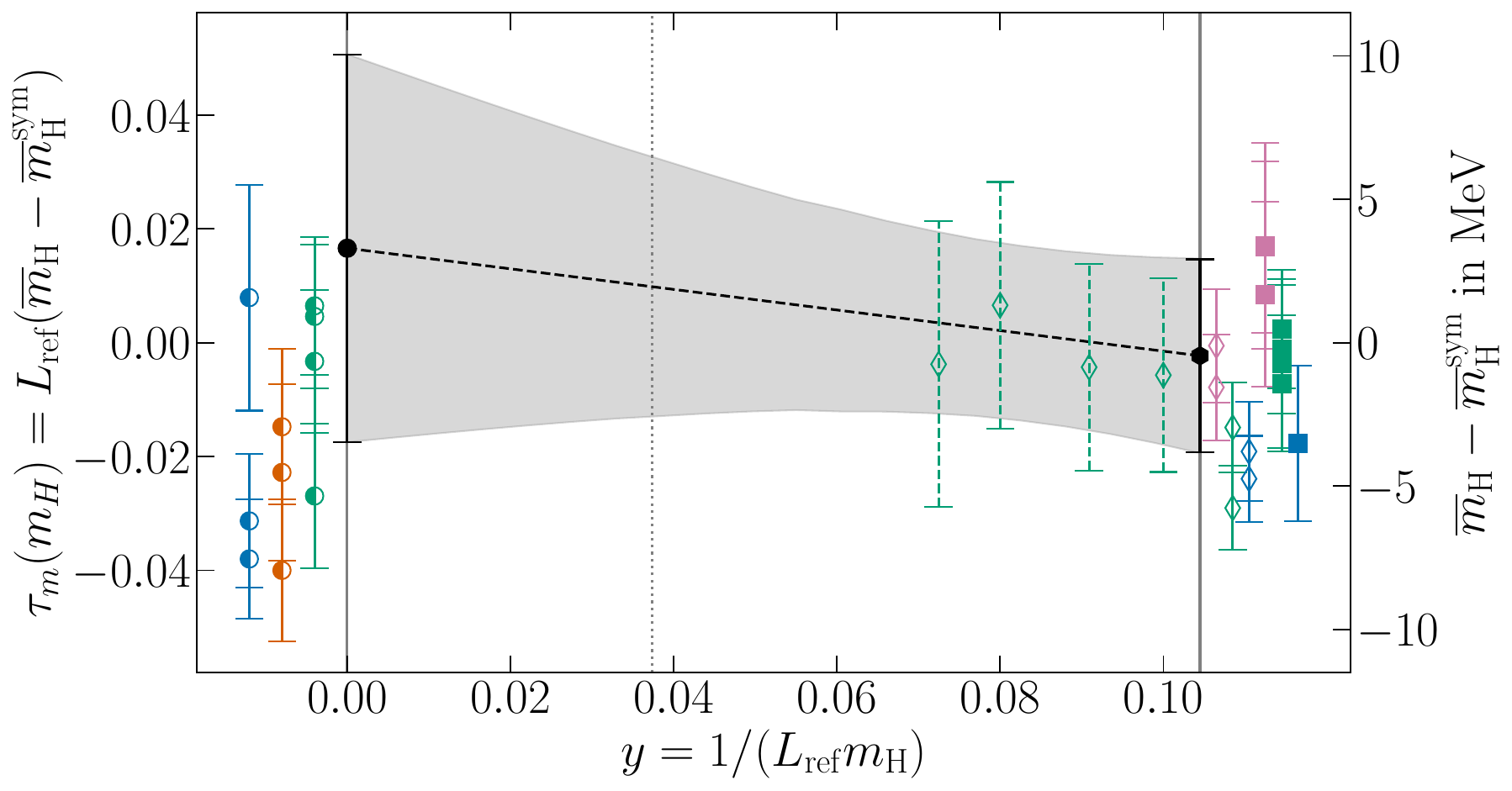}%
	\caption{
		Determination of $\tau_m(m_\pi,m_K)$.
		Left: chiral–continuum extrapolation in the
		static limit.
		Right: interpolation between the static limit and relativistic
		heavy-quark data to reach the bottom-quark mass.
	}
	\label{fig:splitting}
\end{figure}

The determination of $\tau_m(m_\pi,m_K)$ is illustrated in
Figure~\ref{fig:splitting}.
The left panel shows the chiral–continuum extrapolation in the static limit, based on a determination of the
static meson mass at three lattice spacings down to the physical
light-quark mass~\cite{Gerardin:2021jch}.
The physical mass is indicated by the vertical dotted line.
The data are described by the fit ansatz
\begin{align}
	\tau_m^{\rm stat}(m_\pi,m_K, a)
	= L_{\rm ref}^4 (m_K^2 - m_\pi^2)^2 \bigl(c_0 + c_1 a^2\bigr)\,,
\end{align}
with fit parameters $c_0$ and $c_1$.
The colored curves show the fit evaluated at the lattice spacings
entering the analysis, while the gray band represents the continuum
limit.

We perform a corresponding determination at the mass of the charm quark
using the CLS data sets employed for the
charm-quark mass determination with $\mathrm{O}(a)$ improved Wilson
quarks~\cite{Heitger:2021apz} and twisted-mass fermions
~\cite{Bussone:2023kag}, in a combined continuum
extrapolation.

The right panel shows the interpolation between the static limit and
the relativistic result at the charm mass.
The resulting curve is used to determine $\tau_m$ at the bottom-quark
mass indicated by the dotted vertical line.
The colored points to the left and right of the main panel correspond
to the lattice data entering the two chiral–continuum extrapolations
and are horizontally offset for clarity.
The dashed error bars denote additional results at finite lattice
spacing that are not included in the fit but illustrate the heavy-quark mass dependence.
The resulting light-quark mass dependence is negligible for the heavy-quark masses considered.

\section{Results and Outlook}

\begin{figure}[t]
	\centering
		\includegraphics[width=0.48\textwidth]{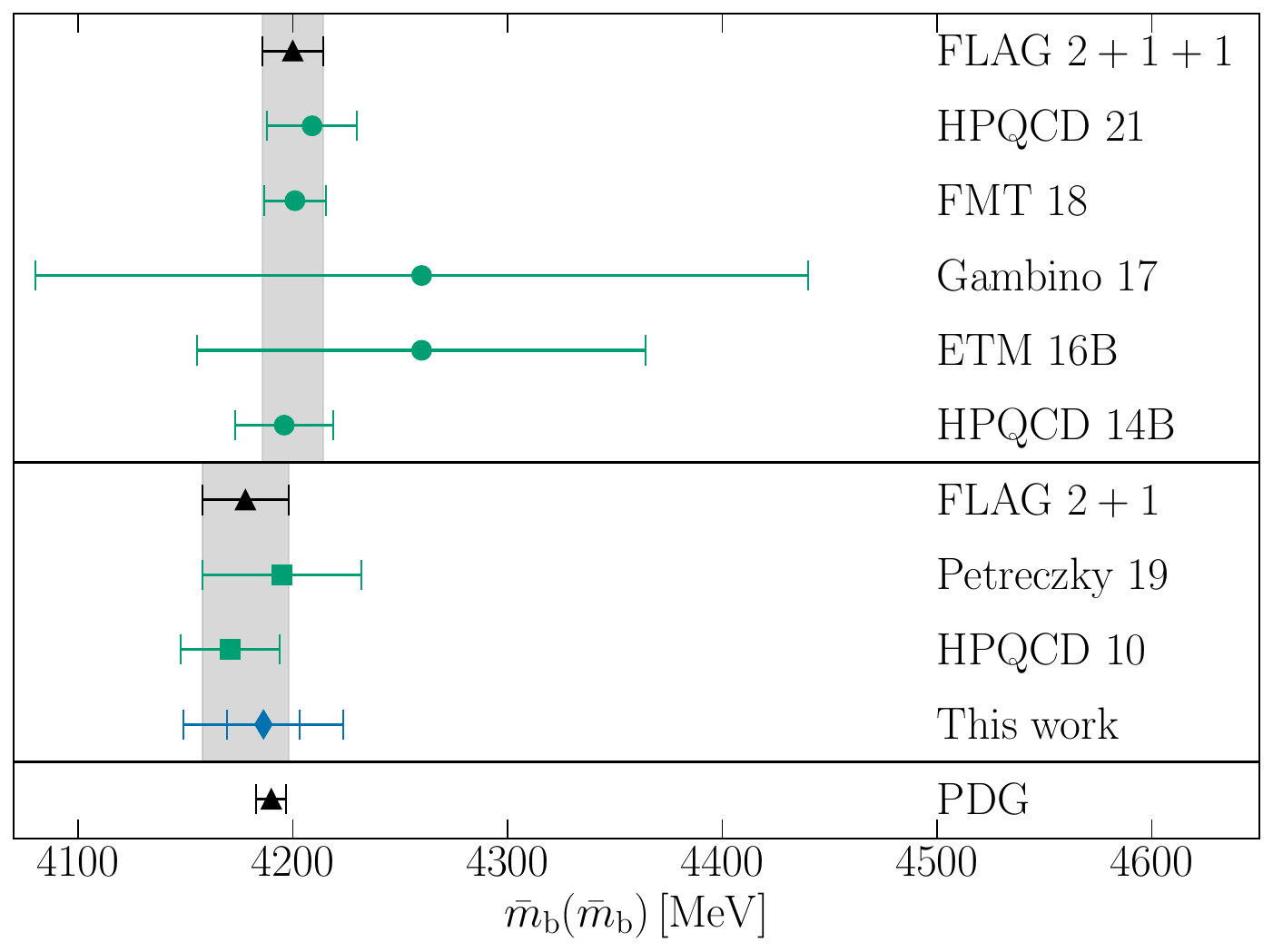}
	\includegraphics[width=0.48\textwidth]{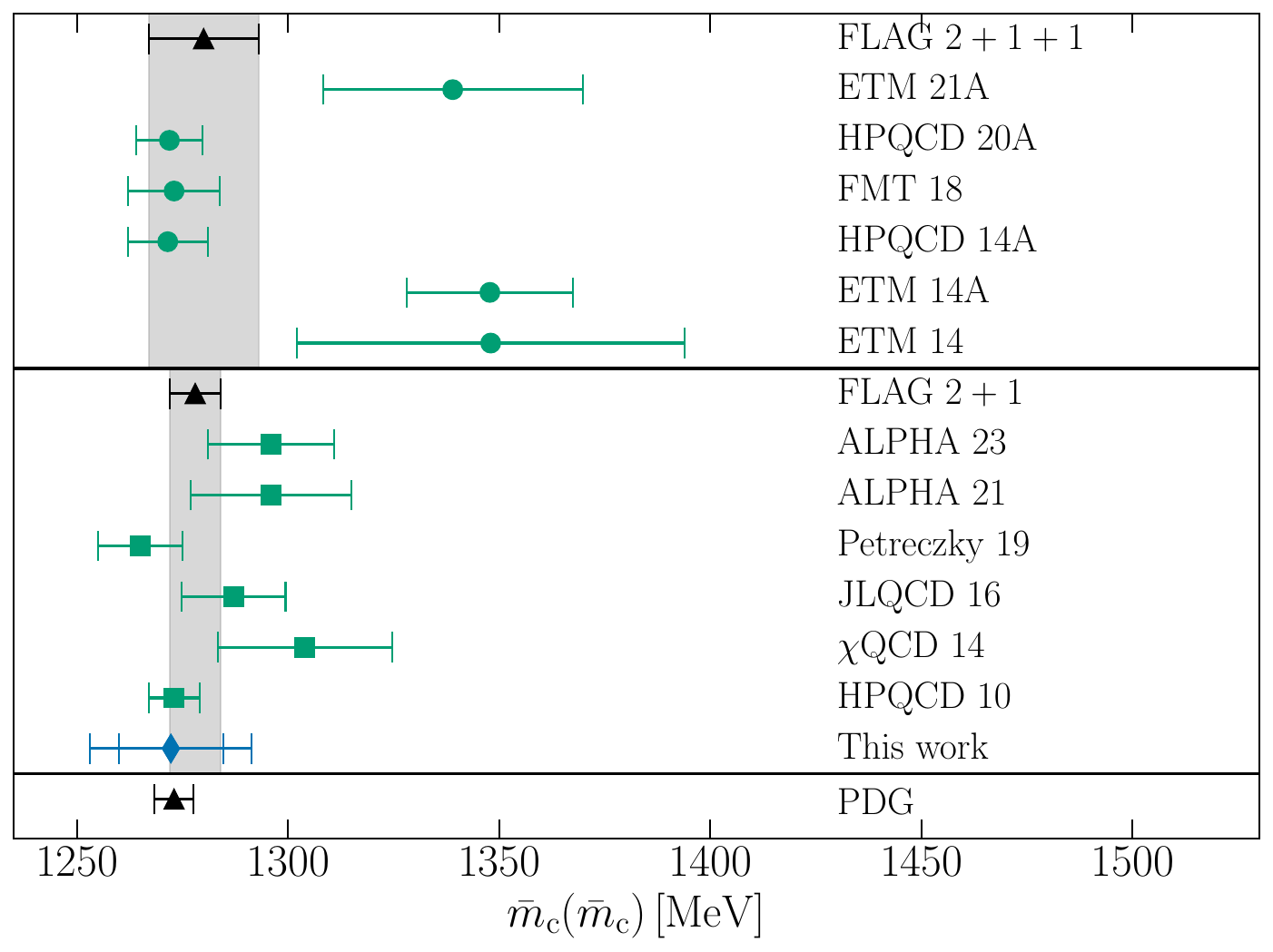}
	\caption{
		Preliminary determinations of the bottom and charm quark masses
		obtained in this work compared with existing results entering the
		FLAG averages~\cite{FlavourLatticeAveragingGroupFLAG:2024oxs,
			Hatton:2021syc,FermilabLattice:2018est,Gambino:2017vkx,
			ETM:2016nbo,Colquhoun:2014ica,Petreczky:2019ozv,
			McNeile:2010ji,ExtendedTwistedMass:2021gbo,Hatton:2020qhk,Chakraborty:2014aca,
			Alexandrou:2014sha,Carrasco:2014cwa,Bussone:2023kag,Heitger:2021apz,Petreczky:2019ozv,
			Nakayama:2016atf,Yang:2014sea}.
		All results are shown in the $\overline{\mathrm{MS}}$ scheme in the
		four-flavor theory.
	}
	\label{fig:results}
\end{figure}

Combining the small-volume determination of the renormalization-group
invariant mass with the nonperturbative step-scaling functions
according to Eq.~(\ref{Lmh_def}) yields the heavy-quark masses.
Our preliminary results are compared with existing determinations in
Figure~\ref{fig:results}.

The outer error bars denote the full statistical and systematic
uncertainty, while the inner error bars exclude the contribution from
the renormalization-group running between the scale $1/L_0$, where the
mass is renormalized, and the RGI mass, as well as the subsequent
conversion to the $\overline{\text{MS}}$ scheme in the four-flavor
theory at a given scale.
This non-perturbative running is taken from~\cite{Campos:2018ahf}.
An improvement of this external input, for example by adding finer
lattices or by using the decoupling strategy
of~\cite{DallaBrida:2019mqg}, could therefore lead
to a significant reduction of the total uncertainty.

The uncertainties are dominated by statistical rather than
systematic effects.
At the present stage the achieved precision is comparable to that of
other $2+1$ flavor determinations.

To finalize the results we are investigating subleading sources of
uncertainty, in particular small mistunings in the line of constant
physics of the Schrödinger functional simulations and the choice of
kinematics in the small-volume setup.
These effects are currently included as a conservative systematic
uncertainty, which is expected to decrease with a more detailed
analysis.

In addition to the quark-mass step-scaling functions we have also
computed those for the axial-vector and vector currents~\cite{Conigli:2023rod,Conigli:2024lxq}.
This will allow the step-scaling strategy to be applied to the
determination of semileptonic $B$-meson decay form factors~\cite{DAnna:2026xgh}, which play an important role in tests of the Standard Model in the heavy-flavor
sector.
This development opens the door to precise lattice predictions for
heavy-flavor observables based on the same step-scaling framework.

\begin{small}
	\section*{Acknowledgments}
	\noindent
	We thank Antonino D'Anna, Oliver Bär and Nikolai Husung for helpful discussions.
	We acknowledge support from the EU projects EuroPLEx
	H2020-MSCAITN-2018-813942 (under grant agreement No.~813942),
	STRONG-2020 (No.~824093) and HiCoLat (No.~101106243),
	as well as from grants,
	PID2024-160152NB-I00, the Momentum CSIC Programme, PID2021-127526NB-I00, SEV-2016-0597,
	CEX2020-001007-S funded by MCIN/AEI,
	the Excellence Initiative of Aix-Marseille University - A*Midex
	(AMX-18-ACE-005),
	and by DFG, through the former
	Research Training Group {GRK 2149} and the project
	``Rethinking Quantum Field Theory'' (No. 417533893/GRK2575).
	The authors gratefully acknowledge the Gauss Centre for Supercomputing e.V. (\url{www.gauss-centre.eu}) for funding this project by providing computing time on the GCS Supercomputer SuperMUC-NG at Leibniz Supercomputing Centre (\url{www.lrz.de}) and the computing time on the high-performance computer "Lise" at the NHR center NHR@ZIB. This center is jointly supported by the Federal Ministry of Education and Research and the state governments participating in the NHR (\url{www.nhr-verein.de}). We furthermore acknowledge the computer resources provided by the CIT of the University of Münster (PALMA-II HPC cluster) and by DESY Zeuthen (PAX cluster) and thank the staff of the computing centers for their support. We are grateful to our colleagues in the CLS initiative for sharing ensembles.
	The determination and propagation of statistical uncertainties is
	performed using the $\Gamma$-method in the implementation of the
	\texttt{pyerrors} package
	\cite{Wolff:2003sm,Ramos:2018vgu,Joswig:2022qfe}. 
\end{small}

\providecommand{\href}[2]{#2}\begingroup\raggedright\endgroup

\end{document}